\documentclass[a4paper,12pt]{article}

\usepackage{amsthm}
\usepackage{amsbsy}
\usepackage{epic}
\usepackage{graphicx}
\usepackage{graphics}
\usepackage{amssymb}
\usepackage{psfrag}
\usepackage{amsmath}
\usepackage{epsfig}
\usepackage{array}
\usepackage[round]{natbib}

\newtheorem{theorem}{Theorem}

\theoremstyle{definition}

\def\N{\mbox{N}}

\def\z{\mathbb{Z}^{(t,x)}}
\def\prob{\mathrm{P}}

\def\expect{\mathbb{E}}

\def\real{\mathbf{R}} 

\def\X{{\bf X}}
\def\Z{{\bf Z}}
\def\W{{\bf W}}
\def\B{{\bf B}}
\def\u{{\bf u}}

\def\z{{\bf z}}
\def\x{{\bf x}}
\def\y{{\bf y}}
\def\bzero{{\bf 0}}

\def\beeta{\mbox{\boldmath$\eta $}}
\def\balpha{\mbox{\boldmath$\alpha $}}
\def\bbeta{\mbox{\boldmath$b $}}
\def\bsigma{\mbox{\boldmath$\Sigma $}}

\def\bI{\mbox{\boldmath$I $}}

\def\low{L_\W}
\def\up{U_\W}

\def\aux{{\bf V}}
\def\auxval{{\bf v}}

\newcommand{\comment}[1]{}

\newcommand{\trans}[3]{\ensuremath{p_{#1}(#3\mid#2)}}

\newcommand{\norm}[2]{\ensuremath{\mathcal{N}_{#1}(#2)}}

\long\def\symbolfootnote[#1]#2{\begingroup%
\def\thefootnote{\fnsymbol{footnote}}\footnote[#1]{#2}\endgroup}

\begin{document}

\title{Particle Filters for Partially Observed Diffusions}
\author{Paul  Fearnhead\thanks{Department of
    Mathematics and Statistics,
    Lancaster University, U.K., email: \texttt{p.fearnhead@lancaster.ac.uk}
}, Omiros Papaspiliopoulos\thanks{Department of
Statistics, Warwick University, U.K., email:
\texttt{O.Papaspiliopoulos@warwick.ac.uk,
 Gareth.O.Roberts@warwick.ac.uk}
\newline
The authors acknowledge helpful comments from the editor and referees.
} and
  Gareth O. Roberts$^{\dagger }$
  }

\maketitle

\begin{abstract} In this paper we introduce a novel particle filter scheme for a class of
partially-observed multivariate diffusions.
We consider a
variety of observation schemes, including diffusion observed with
error, observation of a subset of the components of the
multivariate diffusion and arrival times of a Poisson process
whose intensity is a known function of the diffusion (Cox
process). Unlike currently available methods, our particle filters do not
require approximations of the transition and/or the observation
density using time-discretisations. Instead, they build on recent
methodology for the exact simulation of the diffusion process and the
unbiased estimation of the transition density as described in
\cite{besk:papa:robe:fear:2006}. 
We introduce the Generalised Poisson Estimator, which
generalises the  Poisson Estimator of \cite{besk:papa:robe:fear:2006}.
A central limit theorem is given for our particle filter scheme.
\end{abstract}


\section{Introduction}
\label{sec:intro}

There is considerable interest in using diffusion processes to
model continuous-time phenomena in many diverse scientific disciplines.
These processes can be used to model directly the observed data
 and/or to describe unobserved processes
in a hierarchical model. \comment{The celebrated stochastic
volatility model in financial econometrics is an example where
diffusions are used to model both the observed data (the log-price
of an asset) and unobserved processes (the volatility).}
 This paper focuses on estimating the path of the diffusion given partial information about it.
 We develop
novel particle filters for analysing a class of multivariate diffusions
which are partially observed at a set of discrete time-points.

Particle filtering methods are standard Monte-Carlo methods for
analysing partially-observed discrete-time dynamic models
\cite[]{douc:frei:gord:2001}. They involve estimating the filtering
densities of interest by a swarm of weighted particles. The
approximation error decreases as the number of particles, $N$,
increases. However, filtering for diffusion processes is
significantly harder than for discrete-time Markov models since
the transition density of the diffusion is unavailable in all but
a few special cases. In many contexts even the observation density
is intractable.
Therefore, the standard propagation/weighting/re-sampling steps in
the particle filter algorithm cannot be routinely applied.

To circumvent these complications,  a further approximation, based on a
time-discretisation of the diffusion, has
been suggested \citep[see for
example][]{Crisan/DelMoral/Lyons:1999,mora:jaco:prot:2001}.  The
propagation of each particle from one observation time to the next
is done by splitting the time increment into $M$, say, pieces and
performing $M$ intermediate simulations according to an
appropriate Gaussian distribution. As $M$ gets
large this Gaussian approximation converges to the true diffusion
dynamics. In this framework the computational cost of
the algorithm is of order $M\times N$, and the true filtering
distributions are obtained as both $M$ and $N$ increase.

  Our approach does not rely on time-discretisation, but builds on recent work on
the Exact Algorithm for the simulation of diffusions
\citep{besk:robe:2004,besk:papa:robe:2004,besk:papa:robe:fact} and
on  the unbiased estimation of the diffusion transition density
\citep{besk:papa:robe:fear:2006,besk:papa:robe:smooth}.
 This algorithm  can be used in a
variety of ways to avoid time discretisations in the filtering
problem. The potential of the Exact Algorithm in the filtering
problem was brought up in the discussion of
\cite{besk:papa:robe:fear:2006}, see the
contributions by Chopin, K\"unsch, and in particular
 Rousset and Doucet who
also suggest the use of a random weight particle filter
in this context.

One possibility is simply to use the Exact Algorithm  to propagate
the particles in the implementation of the
\cite{Gordon/Salmond/Smith:1993} bootstrap particle filter, thus
avoiding entirely the $M$ intermediate approximate simulations
between each pair of observation times. We call this the Exact
Propagation Particle Filter (EPPF). Where possible, a better approach is to adapt
the Exact Algorithm to simulate directly from (a particle
approximation to) the filtering density using rejection sampling;
we term this the Exact Simulation Particle Filter (ESPF).

However, our favoured method goes in a different direction. We
work in the framework of the auxiliary particle filter of
\cite{pitt:shep:1999}, where particles are propagated from each
observation time to the next according to a user-specified density
and then are appropriately weighted to provide a consistent
estimator of the new filtering distribution. Due to  the
transition density being unavailable, the weights associated with
each particle are  intractable. However, our approach is to assign
to each particle a random positive weight which is an unbiased
estimator of the true weight. We call this the Random Weight Particle Filter (RWPF).
Our algorithm yields consistent estimates of the filtering
distributions. The replacement of the weights in a particle filter
by positive unbiased estimators is an interesting possibility in
more general contexts than the one considered in this paper.
Indeed, in Section \ref{sec:aux} we show that this approach
amounts to a convenient augmentation of the state with auxiliary
variables.

The construction of the unbiased estimators of the weights is one
of the main contributions of this paper, and it is of independent
interest.
This is based on an extension of the {\em Poisson Estimator} of
\cite{besk:papa:robe:fear:2006}, which we call the {\em
Generalised Poisson Estimator}.  This estimator
is guaranteed to return positive estimates (unlike the Poisson
Estimator) and its efficiency (in terms of variance and
computational cost) can be up to orders of magnitude better than
the Poisson Estimator. Optimal implementation of the Poisson and
the Generalised Poisson estimators is thoroughly investigated
theoretically and via simulation.

All three time-discretisation-free particle filters we introduce
are easy to implement, with the RWPF being the easiest and the
most flexible to adapt to contexts more general than those
considered here. A simulation study is carried out which shows
that the RWPF is considerably more efficient than the ESPF which
is more efficient  than the EPPF. We also
provide a theoretical result which shows that our filters can have
significant computational advantages over time-discretisation
methods. We establish  a Central Limit Theorem (CLT) for the
estimation of expectations of the filtering distributions using
either of the EPPF, ESPF and the RWPF. This is an extension of the
results of \cite{Chopin:2004}. The CLT shows that,
for a fixed computational cost $K$, the errors in the particle
approximation of the filtering distributions decrease as
$K^{-1/2}$ in our methods, whereas it is known that the rate is
$K^{-1/3}$ or slower in time-discretisation methods.

The main limitation of the methodology presented here is the requirement
that the stochastic differential equation specifying the underlying
diffusion process can be transformed to one with orthogonal diffusion
matrix, and gradient drift.  Although this framework excludes
some important model types (such as stochastic volatility models)
it incorporates a wide range of processes which can model successfully
many physical processes. On the other hand, our methods can handle a
variety of discrete-time observation schemes. In this paper we consider
three schemes: noisy observations of a diffusion process, observation
of a subset of the components of a multivariate diffusion, and arrival
times of a Poisson process whose intensity is stochastic and it is given
by a known function of a diffusion.

The paper is organised as follows. Section \ref{sec:models}
introduces the model for the underlying diffusion and the
necessary notation, the observation schemes we consider and the
simulated data sets on which we test our proposed methods. Section
\ref{PF} introduces the RWPF and states the CLT. Section
\ref{sec:GPE} introduces the main tool required in constructing
the RWPF, the Generalised Poisson Estimator (GPE).
Several theoretical results
are established for the GPE, and
a simulation study is performed to assess its performance. Section
\ref{sec:comp} is devoted in the empirical investigation of the
performance of the different particle filters we introduce.  Several
implementation issues are also discussed.  Section \ref{sec:discuss}
closes with a discussion on  extensions of the methodology and the
appendices contain technical results and proofs.

\section{Signal, data and assumptions}
\label{sec:models}

%
Let the signal be modelled by a $d$-dimensional diffusion process
\begin{equation}
\label{eq:SDEmult} \mathrm{d}\X_s = \balpha(\X_s)\, \mathrm{d}s +
 \mathrm{d}\B_s\ , \quad s \in [0,t]\,.
\end{equation}
\noindent We assume throughout the paper that the drift
is known.
Our approach requires some assumptions which we summarize in this
paragraph:
  i) $\balpha$ is continuously differentiable in all its
arguments, ii) there exists a function $A:{\mathbf R}^d \to
{\mathbf R}$ such that $\balpha (\u) = \nabla A (\u)$, and iii)
there exists $l > -\infty $ such that $\phi (\u ) :=
\bigl(\|\balpha (\u) \|^2 + \nabla^2 A (\u)\bigr)/2 - l \ge 0$.
\label{pg:phi} Among these last three conditions i) and iii) are
weak and the strictest is ii), which in the ergodic case
corresponds to $\X$ being a time-reversible diffusion.

\comment{To simplify the notation we will present our methodology
assuming that the original process $\X$ solves an SDE with unit
diffusion coefficient as in (\ref{eq:unitSDEmult}). The extension
of the methodology to the general case simply involves replacing
appropriately $\X$ with $\beeta(\X)$ in some expressions. We will
return to this point in Section \ref{sec:disc}. The examples we
consider }

The transition density of (\ref{eq:SDEmult}) is typically
intractable but 
a useful expression
is available \citep[see for
example][]{besk:papa:robe:fear:2006,dacu:flor:1986}
\begin{equation} \label{eq:td}
\trans{t}{\x_0}{\x_t}=\norm{t}{\x_t-\x_0}\exp\{A(\x_t)-A(\x_0) -lt
\}\expect\left [ \exp \left \{-\int_0^t\phi(\W_s)\mbox{d}s \right
\}\right]\,.
\end{equation}
In this expression $\norm{t}{\u}$ denotes the density of the
$d$-dimensional normal distribution with mean $\bzero$ and
variance ${t}\bI_d$ evaluated at ${\u}\in\mathbf{R}^d$, and the
expectation is taken w.r.t.\@ a Brownian bridge, $\W_s, s \in
[0,t]$, with $\W_0=\x_0$ and $\W_t=\x_t$. Note that the
expectation in this formula typically cannot be evaluated.

The data consist of  partial observations $y_1,y_2,\ldots,y_n$, at
discrete time-points $0 \leq t_1 < t_2 <\cdots< t_n $. We consider
three possible observation regimes:
\begin{itemize}
\item[(A)] {\it Diffusion observed with error}. The observation
$y_i$, is related to the signal  at time $t_i$ via a known density
function $f(y_i|\x_{t_i})$. This model extends the general
state-space model by allowing the signal to evolve continuously in
time. There is a wide range of applications which fit in this
framework, see \cite{douc:frei:gord:2001} for references.
\item[(B)] {\it Partial Information}. At time $t_i$ we observe
$y_i=\zeta(\X_{t_i})$ for some non-invertible known function
$\zeta(\cdot)$. For example we may observe a single component of
the $d$-dimensional diffusion.  In this model type
$f(y_i|\x_{t_i})=1$ for all $\x_{t_i}$ for which
$\zeta(\x_{t_i})=y_i$.
 \item[(C)] {\it Cox Process}. In this regime the data consist of
the observation times $t_i$ which are random and are assumed to be
the arrivals of a Poisson process of rate $\nu(\X_s)$, for some
known function $\nu$. Such models are popular in insurance
\citep{dass:jang:2005} and finance
\citep{Engel:2000,Duff:Sing:mode:1999}, and they have recently been
used to analyse data from single molecule experiments
\citep{kou:xie:liu:2005}.
There is a
significant difference between this observation regime and the two
previous ones. To have notation consistent with (A) and (B) we let
$y_i=t_i$ denote the time of the $i$th observation; and define the
likelihood $f(y_i \mid\x_{t_{i-1}},\x_{t_{i}})$ to be the
probability density that the next observation after $t_{i-1}$ is
at time $t_i$. This density can be obtained by integrating
\begin{equation}
\label{eq:cond-int} \nu(\X_s) \exp \left \{-\int_{t_{i-1}}^{t_i}
\nu(\X_s) \mbox{d}s \right \},
\end{equation}
w.r.t.\@ the distribution of $(\X_s,s\in(t_{i-1},t_i))$
conditionally on $\X_{t_{i-1}}=\x_{t_{i-1}}, \X_{t_i}=\x_{t_i}$.
The distribution of this conditioned process has a known density
w.r.t.\@ the Brownian bridge measure and it is given in Lemma 1 of
\cite{besk:papa:robe:fear:2006}. We can thus show that the density
of interest is
\begin{equation}
\label{eq:cox-lik} \frac{\nu(\x_{t_i})
\norm{t_i-t_{i-1}}{\x_{t_i}-\x_{t_{i-1}}}}{\trans{t_i-t_{i-1}}{\x_{t_{i-1}}}{\x_{t_i}}}
\exp\{A(\x_{t_i})-A(\x_{t_{i-1}})\}\expect\left [ \exp \left
\{-\int_{t_{i-1}}^{t_i}(\phi(\W_s)+\nu(\W_s))\mbox{d}s \right
\}\right]\,,
\end{equation}
where expectation is with respect to the law of a Brownian Bridge
from $\x_{t_{i-1}}$ to $\x_{t_i}$. 
\end{itemize}

We take a Bayesian approach, and assume a prior distribution for
$\X_0$. Our interest lies in the online calculation of the
filtering densities, the posterior densities of the signal at time
$t_i$ given the observations up to time $t_i$, for each $1 \leq i
\leq n$. While these densities are intractable, we propose a
particle filter scheme to estimate recursively
these densities at each observation time-point. As we
point out in Section 6, our approach allows the estimation of the
filtering distribution of the continuous time path
$(\X_s,t_{i-1}<s<t_i)$.

A more flexible model for the signal is a diffusion process $\Z$
which solves a more general SDE than the one we have assumed in
(\ref{eq:SDEmult}): 
\begin{equation}
\label{eq:SDEmultGen} \mathrm{d}\Z_s = \bbeta(\Z_s)\, \mathrm{d}s
+
 \bsigma(\Z_s) \mathrm{d}\B_s\ , \quad s \in [0,t]\,.
\end{equation}
In contrast with (\ref{eq:SDEmult}), (\ref{eq:SDEmultGen}) allows
the diffusion coefficient to be state-dependent. Our methods
directly apply to all such processes provided there is an
explicit transformation $\Z_s \mapsto \beeta(\Z_s)=:\X_s$, where
$\X$ solves an SDE of the type (\ref{eq:SDEmult}); the implied
drift $\balpha$ can be easily expressed in terms of $\bbeta$ and
$\bsigma$ via It\^o's formula and it will have  to satisfy the
conditions we have already specified. In model
(A) the likelihood becomes $f(y_i \mid \beeta^{-1}(\X_{t_i}))$, in
model (B) the data are $y_i = \zeta(\beeta^{-1}(\X_{t_i}))$ and in
model (C) the Poisson intensity is $\nu(\beeta^{-1}(\X_s))$, where
$\beeta^{-1}$ denotes the inverse transformation. Therefore, the
extension of our methodology to general diffusions is
straightforward when $d=1$; under mild conditions
(\ref{eq:SDEmultGen}) can be transformed to (\ref{eq:SDEmult}) by
$\eta(Z_s)= \int_{u^{*}}^{Z_s}\Sigma(z)^{-1}\, \mathrm{d}z$, for
some arbitrary $u^{*}$ in the state space of the diffusion.
Moreover, the drift of the transformed process will typically
satisfy the three conditions we have specified. However, the
extension is harder in higher dimensions. The necessary
transformation is more complicated when $d>1$ and it might be
intractable or even impossible \citep{saha:multi}. Even when such
a transformation is explicit it might imply a drift for $\X$ which
violates condition ii). Nevertheless, many physical
systems can be successfully modeled with diffusions which can be
transformed to (\ref{eq:SDEmult}).

Our particle filtering methods will be illustrated on two sets of
simulated data:

\noindent{\bf Example 1: Sine diffusion observed with error}. The
signal satisfies
\begin{equation}
\label{eq:sine-sde} \mbox{d} X_s =  \sin(X_s)\mbox{d}s
+\mbox{d}B_s\,,
\end{equation}
and the data consist of noisy observations, $y_i \sim
\N(X_{t_i},\sigma^2)$. Figure \ref{Fig:1}(top) shows a simulation
of this model with
$\sigma=0.2$. In this case
\begin{equation}\label{eq:phi}\phi(u) = (\sin(u)^2+\cos(u)+1)/2\,.\end{equation}
 This process is closely related to Brownian motion on a circle.
It is convenient as an
illustrative example since discrete-time skeletons can be easily
simulated from this process using the most basic form of the Exact
Algorithm \citep[EA1 in][R-code is available on
request by the authors]{besk:papa:robe:2004}.

\noindent{\bf Example 2: OU-driven Cox Process}. The second data
set consists of the arrival times of a Poisson process,
 $y_i = t_i,$
whose intensity is given by $\nu(X_s),s\geq 0$, where
$$\nu(x) = a + \beta |x|,$$
and $X$ is an Ornstein-Uhlenbeck (OU) process,
$$\mbox{d} X_s = -\rho X_s \mbox{d}s +
\mbox{d}B_s\,.$$
The OU process is stationary with Gaussian marginal distribution,
$\N(0, 1/(2 \rho))$. Thus, an interpretation for this model is
that the excursions of $X$ increase the Poisson intensity, whereas
$a$ corresponds to the intensity when $X$ is at its mean level. An
example data set is shown in Figure \ref{Fig:3}; where
 we have taken $a=0$, $\beta=20$, $\rho=1/2$.
 Although the transition
density of the OU process is well-known,
$$X_t \mid X_0=x_0 \sim \N \left (e^{-\rho t}x_0,\frac{1}{2 \rho} (1-e^{-2\rho t})
\right )\,,$$
the observation density $f(y_{i+1} \mid x_{t_i},x_{t_{i+1}})$ is
intractable.
%
%

Examples 1 and 2 are examples of observation regimes (A) and (C)
respectively.  We will show that observation regime (B) can be handled in
a similar fashion as (A), so we have not included an accompanying example.


\section{Random weight particle filter}\label{PF}

As in Section \ref{sec:models} we will denote the observation at
time $t_i$ by $y_i$, and $p_t(\cdot\mid \cdot)$ will denote the
system transition density over time $t$ (see Equation
\ref{eq:td}). We will write $\Delta_i = t_{i+1}-t_i$, and the
filtering densities $p(\x_{t_i}|y_{1:i})$ will be denoted by
$\pi_i(\x_{t_i})$, where by standard convention
$y_{1:i}=(y_1,\ldots,y_i)$. To simplify notation, when we
introduce weighted particles below, we will subscript both
particles and weights by $i$ rather than $t_i$.

Our aim is to recursively calculate the filtering densities
$\pi_i(\x_{t_i})$. Basic probability calculations yield the following
standard filtering recursion for these densities
\begin{equation} \label{eq:1}
\pi_{i+1}(\x_{t_{i+1}}) \propto \int
f(y_{i+1}|\x_{t_i},\x_{t_{i+1}})p_{\Delta_i}(\x_{t_{i+1}}|\x_{t_i})
\pi_i(\x_{t_i})\mbox{d}\x_{t_i}.
\end{equation}
Particle filters approximate $\pi_i(\x_{t_i})$  by a discrete
distribution, denoted by $\hat{\pi}_i(\x_{t_i})$, whose support is a
set of $N$ particles, $\{\x_{i}^{(j)}\}_{j=1}^N$,
with associated probability  weight $\{w_i^{(j)}\}_{j=1}^N$.
  Substituting $\hat{\pi}_i(\x_{t_i})$ for
$\pi_i(\x_{t_i})$ in (\ref{eq:1}), yields a (continuous density)
approximation to $\pi_{i+1}(\x_{t_{i+1}})$,
\begin{equation} \label{eq:pitilde}
\tilde{\pi}_{i+1}(\x_{t_{i+1}}) \propto \sum_{j=1}^N w_i^{(j)}
f(y_{i+1}|\x^{(j)}_{i},\x_{t_{i+1}})p_{\Delta_i}(\x_{t_{i+1}}|\x^{(j)}_{i}).
\end{equation}
The aim of one iteration of the particle filter algorithm is to
construct a further particle (discrete distribution) approximation
to $\tilde{\pi}_{i+1}(\x_{t_{i+1}})$.

We can obtain such a  particle approximation  via importance
sampling, and a general framework for achieving this is given by
the auxiliary particle  filter of \cite{pitt:shep:1999}. We choose
a proposal density of the form
\begin{equation}
\label{eq:proposal}
\sum_{j=1}^N \beta_i^{(j)}
q(\x_{t_{i+1}}|\x^{(j)}_i,y_{t_{i+1}})\,. \end{equation}
Choice of suitable proposals, i.e.\@ choice of the
$\beta_i^{(j)}$s and $q$, is discussed in the analysis of our
specific applications in Section \ref{sec:comp}. 

To simulate a new particle at time $t_{i+1}$ we (a) simulate a
particle $\x_{i}^{(k)}$ at time $i$, where $k$ is a realisation of
a discrete random variable which takes the value $j \in
\{1,2,\ldots,N\}$ with probability $\beta_i^{(j)}$; and (b)
simulate a new particle at time $t_{i+1}$ from
$q(\x_{t_{i+1}}|\x^{(k)}_i,y_{i+1})$. The weight assigned to this
pair of particles $(\x_{i}^{(k)},\x_{t_{i+1}})$ is proportional to
\begin{equation} \label{eq:2}
\frac{w^{(k)}_i
f(y_{i+1}|\x^{(k)}_{i},\x_{t_{i+1}})p_{\Delta_i}(\x_{t_{i+1}}|\x^{(k)}_{i})}
{\beta_i^{(k)}q(\x_{t_{i+1}}|\x^{(k)}_i,y_{i+1})}\,.
\end{equation}
This is repeated $N$ times
to produce the set of weighted particles at time
$t_{i+1}$, $\left\{(\x_{{i+1}}^{(j)},w_{{i+1}}^{(j)}) \right
\}_{j=1}^{N}$, which gives an importance sampling approximation to
$\pi_{i+1}(\x_{t_{i+1}})$. Renormalising the weights is possible
but does not materially affect the methodology or its accuracy.
Improvements on independent sampling in
step (a) can be made: see the stratified sampling ideas of
\cite{Carpenter/Clifford/Fearnhead:1999}. The
resulting particle filter has good theoretical properties
including consistency \cite[]{Crisan:2001}
and central limit theorems for estimates
of posterior moments \cite[]{DelMoral/Miclo:2000,Chopin:2004,Kunsch:2005},
as $N\rightarrow\infty$.
Under conditions relating to exponential forgetting of initial conditions,
particle filter errors stabilise as $n\rightarrow\infty$
\cite[]{DelMoral/Guionnet:2001,Kunsch:2005}. 

The difficulty with implementing such a particle filter when the
signal $\X$ is a diffusion process is that the transition density
$p_{\Delta_i}(\x_{t_{i+1}}|\x^{(k)}_{i})$ which appears in
(\ref{eq:2}) is intractable for most diffusions  of interest, due
to the expectation term in (\ref{eq:td}). Furthermore, for
observation model (C) (but also for more general models), the
likelihood term $f(y_{i+1}|\x^{(k)}_{i},\x_{t_{i+1}})$  given in
(\ref{eq:cox-lik}) cannot be calculated analytically.

We circumvent these problems by assigning each new particle a
random weight which is a realisation of a random variable whose
mean is (\ref{eq:2}). The construction and simulation of this
random variable is developed in Section \ref{sec:GPE}, and it is
based on the particular expression for the transition density in
(\ref{eq:td}). The replacement of the weights by positive unbiased
estimators is an interesting possibility in more general contexts
than the one considered in this paper. Indeed, in Section
\ref{sec:aux} we show that this approach amounts to a convenient
augmentation of the state with auxiliary variables.

\subsection{Simulation of weights}
\label{sec:sim-w}

In all models the weight associated with the
pair $(\x_{i}^{(k)},\x_{t_{i+1}})$ equals
\begin{equation} \label{eq:wirghts-prod}
h_{i+1}(\x_{i}^{(k)},\x_{t_{i+1}},y_{i+1})\mu_g(\x_{i}^{(k)},\x_{t_{i+1}},t_i,t_{i+1})
\end{equation}
where $h_{i+1}$ is a known function, and for
$0<u<t$,
\begin{equation*} 
\mu_g(\x,\z,u,t)\,:=\,\expect\left[ \exp \left \{-\int_{u}^{t}
g(\W_s)\mbox{d}s \right \}\right],
\end{equation*}
where the expectation is taken w.r.t.\@ a $d$-dimensional Brownian
bridge $\W$, starting at time $u$ from $\W_{u}=\x$ and finishing
at time $t$ at $\W_t=\z$.
\newline

\noindent {\bf Models (A) and (B)} : For these model types
\[
h_{i+1}(\x_{i}^{(k)},\x_{t_{i+1}},y_{i+1}) = \frac{w^{(k)}_i
f(y_{i+1}|\x_{t_{i+1}})\norm{\Delta_i}{\x_{t_{i+1}}-\x_{i}^{(k)}}
\exp\{A(\x_{t_{i+1}})-A(\x_{i}^{(k)})\}} {\beta_i^{(k)}
q(\x_{t_{i+1}}|\x^{(k)}_i,y_{i+1})}\, , \]
and $g=\phi$.
In
 model type (B)
  the
proposal distribution $q(\x_{t_{i+1}}|\x_{i}^{(k)},y_{i+1})$
should be chosen to  propose only values of $\x_{t_{i+1}}$ such
that $\zeta(\x_{t_{i+1}})=y_{i+1}$; then
$f(y_{i+1}|\x_{t_{i+1}})=1$. \newline

\noindent {\bf Model (C)}: \label{(C)} A synthesis of
(\ref{eq:td}), (\ref{eq:cox-lik}) and (\ref{eq:2}), with
$g=\phi+\nu$ gives
\[
h_{i+1}(\x_{i}^{(k)},\x_{t_{i+1}},y_{i+1}) = \frac{w^{(k)}_i
\nu(\x_{t_{i+1}}) \norm{\Delta_i}{\x_{t_{i+1}}-\x_{i}^{(k)}}
\exp\{A(\x_{t_{i+1}})-A(\x_{i}^{(k)})\}} {\beta_i^{(k)}
q(\x_{t_{i+1}}|\x^{(k)}_i,y_{i+1})}\, , \]

\noindent Section \ref{sec:GPE} shows how to construct for each
pair of $(\x,\z)$ and times $(u,t)$, with $u<t$, additional {\em
auxiliary} variables $\aux$, and a function $r(\auxval,\x,\z,u,t)
\geq 0$, with the property that $\expect[r(\aux,\x,\z,u,t) \mid
\x,\z]=\mu_g(\x,\z,u,t)$. The auxiliary variables are simulated
according to an appropriate conditional distribution $Q_g(\,\cdot
\mid \x,\z,u,t)$, and $r$ is easy to evaluate. Our method replaces
in the weight the intractable term $\mu_g$ with its unbiased
estimator $r$.
\newline

{\em \noindent {\bf Random Weight Particle Filter (RWPF)} \\
\newline PF0  Simulate a sample $\x_0^{(1)},\ldots,\x_0^{(N)}$
from $p(\x_0)$, and set $w_0^{(j)} = 1/N$.
\newline
 For $i=0,\ldots,n-1$, for $j=1,\ldots,N$:
\begin{itemize}
\item[PF1] calculate the effective sample size of the
$\{\beta_i^{(k)}\}$, $ESS =
(\sum_{k=1}^N (\beta_i^{(k)})^2)^{-1}$; if $ESS <C$, for some fixed
constant $C$,
 simulate $k_{i,j}$ from
$p(k)=\beta_{i}^{(k)}$, $k=1, \ldots,N $ and set
$\delta_{i+1}^{(j)}=1$; otherwise set $k_{i,j}=j$ and
$\delta_{i+1}^{(j)}=\beta_{i}^{(j)}$; \item[PF2] simulate
$\x_{i+1}^{(j)}$ from
$q(\x_{t_{i+1}}|\x_{i}^{(k_{i,j})},y_{i+1})$; \item[PF3] simulate
$\auxval_{i+1} \sim Q_g(\,\cdot \mid
\x_{i}^{(k_{i,j})},\x_{t_{i+1}},t_i,t_{i+1})$; \item[PF4] assign
particle $\x_{i+1}^{(j)}$ a weight
\begin{equation} \label{eq:RW}
w_{i+1}^{(j)} =   \delta_{i+1}^{(j)}
h_{i+1}(\x_{i}^{(k_{i,j})},\x_{i+1}^{(j)},y_{i+1})
r(\auxval_{i+1},\x_{i}^{(k_{i,j})},\x_{t_{i+1}},t_i,t_{i+1})\,.
\end{equation}
\end{itemize}
} Notice that this algorithm contains a decision as to whether or
not resample particles before propagation in step PF1, with
decision being based on the ESS of the
$\beta_i^{(j)}$. The constant $C$ can be interpreted as the minimum acceptable
effective sample size. (See Liu and Chen 1998 for the rationale of
basing resampling on such a condition.)  Whether or not resampling
occurs will affect the weight given to the new sets of particles,
and this is accounted for by different values of
$\delta_{i+1}^{(j)}$ in PF1. Optimally, the resampling for step
PF1 will incorporate dependence across the $N$ samples; for
example the stratified sampling scheme of
\cite{Carpenter/Clifford/Fearnhead:1999} or the residual sampling
of \cite{liu:chen:1998}.



\subsection{An equivalent formulation via an augmentation of the state}
\label{sec:aux}

In the previous section we described a generic sequential Monte
Carlo scheme where the exact weights in the importance sampling
approximation of the filtering distributions are replaced by
positive unbiased estimators. We now show that this scheme is
equivalent to applying an ordinary auxiliary particle filter to a
model with richer latent structure. We demonstrate this equivalent
representation for model types (A) and (B), since an obvious
modification of the argument establishes the equivalence for model
type (C).

According to our construction, conditionally on $\X_{t_i}$,
$\X_{t_{i+1}}$, $t_{i}$ and $t_{i+1}$, $\aux_{i+1}$  is
independent of $\aux_j$ and $\X_{t_j}$ for any $j$ different from
$i,i+1$. Additionally, it follows easily from the unbiasedness and
positivity of $r$ that, conditionally on $\X_{t_i}=\x$,
$r(\auxval_{i+1},\x,\x_{t_{i+1}},t_{i},t_{i+1})$ is a probability
density function for $(\X_{t_{i+1}},\aux_{i+1})$ with respect to
the product measure $Leb(\mbox{d} \z) \times Q_g(\mbox{d} {\bf v}
\mid \x,\z,t_i,t_{i+1})$, where $Leb$ denotes the Lebesgue
measure.

Consider now an alternative discrete-time model with unobserved
states $(\Z_i,\aux_i),i=1,\ldots,n$, $\Z_i \in {\mathbf R}^d$,
with a non-homogeneous Markov transition density
%
$$p_{i+1}(\z_{{i+1}},\auxval_{i+1} \mid \z_{i},\auxval_i) =
r(\auxval_{i+1},\z_{i},\z_{i+1},t_i,t_{i+1})\,,$$
(this density is with respect to $Leb \times Q_g$) and observed
data $y_i$ with observation density $f(y_{i+1} \mid
\z_i,\z_{i+1})$. By construction the marginal
filtering distributions of $\Z_i$ in this model are precisely
$\pi_i(\x_{t_i})$, i.e.\@ the filtering densities in (\ref{eq:1}).
Consider an auxiliary particle filter applied to this model where
we choose with probability $\beta_i^{(j)}$ each of the existing
particles $(\z_{i}^{(j)},\auxval_i^{(j)})$, and generate new
particles according to the following proposal 
$$
(\z_{i+1},\auxval_{i+1}) \sim q(\z_{i+1} \mid  \z_i^{(k)},y_{i+1})
Q_g(\mbox{d} \auxval_{i+1} \mid \z_i^{(k)},\z_{i+1},t_i,t_{i+1})
Leb(\mbox{d} \z_{i+1})\,,
$$
where $q$ is the same proposal density as in (\ref{eq:proposal}).
The weights associated with each particle in this discrete-time
model are tractable and are given  by (\ref{eq:RW}). Therefore,
the weighted sample $\left\{(\z_{{i+1}}^{(j)},w_{{i+1}}^{(j)})
\right \}_{j=1}^{N}$ is precisely a particle approximation to
$\pi_{i+1}(\x_{t_{i+1}})$, and RWPF is equivalent to an auxiliary
particle filter on this discrete-time model whose latent structure
has been augmented with the auxiliary variables $\aux_i$.

This equivalent representation sheds light on many aspects of our
method. Firstly, it makes it obvious that it is inefficient to
average more than one realization of the positive unbiased
estimator of $\mu_g$ per particle. Instead it is more efficient to
generate more particles with only one realization of the estimator
simulated for each pair of particles.

Secondly, it illustrates that RWPF combines the advantages of the
bootstrap and the auxiliary particle filter. Although it is easy
to simulate from the probability distribution $Q_g$ (as described
in Section \ref{sec:GPE}), it is very difficult to derive its
density.
 (with respect to an appropriate reference measure).
Since
the $\aux_i$s are propagated according to this measure, its
calculation is avoided. This is an appealing feature of the
bootstrap filter which propagates particles without requiring
analytically the system transition density. On the other hand the
propagation of the $\Z_i$s is done via a user-specified density
which incorporates the information in the data.

Thirdly, it suggests that the RWPF will have similar theoretical
properties with auxiliary particle filters applied to
discrete-time models. This is explored in Section \ref{sec:clt}.

\comment{
\subsection{Alternative particle filters}

The Exact Algorithm
for diffusion simulation
can be used to
construct alternative filtering algorithms to the Random Weight
Particle Filter (RWPF). 
The {\em Exact Propagation Particle Filter} (EPPF), is a
Bayesian bootstrap filter \cite[]{Gordon/Salmond/Smith:1993} where
the Exact Algorithm is used to propagate particles from the system
SDE. Alternatively the {\em Exact Sampling
Particle Filter} (ESPF), uses the Exact Algorithm to perform
rejection sampling from (\ref{eq:pitilde}) in the spirit of
\cite{hurz:kuns:1998}; see also the discussion in
\cite{besk:papa:robe:fear:2006}, and \cite{Kunsch:2005} for more
detailed theoretical properties of this general rejection sampling
approach to filtering. A comparison of the different approaches is
given in Section \ref{sec:comp} where we argue why we favour the
RWPF.

}

\subsection{Theoretical properties}
\label{sec:clt}

Consider estimation of the posterior mean of some function
$\varphi$ of the state at time $t_i$,
$\expect[\varphi(\x_{t_i})|y_{1:i}]$. A natural approach to the
investigation of particle filter effectiveness is to consider
the limiting behaviour of the algorithm as $N \to \infty $. For
the standard auxiliary particle filter, \cite{Chopin:2004}
introduces a central limit
theorem (CLT) for estimation of this type of expectations. 
This CLT applies directly to both EPPF and the ESPF.

In Appendix F  we extend the result of \cite{Chopin:2004} and give
a further necessary condition on the random weights in  RWPF under
which a CLT still holds. This extra condition is (C2). The
expression for the variance of the estimator of
$\expect[\varphi(\x_{t_i})|y_{1:i}]$ obtained with  RWPF differs
from the expression in the standard case (i.e.\@ when the weights
are known) by an extra term caused by the randomness in the
weights (see Equations \ref{CLT1}--\ref{CLT3} and the comment on
Theorem \ref{thm:CLT} in Appendix F for further details). The
ready adaptation of Chopin's approach is facilitated by the
observation that the RWPF can be re-expressed as a standard
particle filter for the an augmented state (see Section
\ref{sec:aux}).

One important consequence of this CLT is that the errors in
estimating $\expect[\varphi(\x_{t_i})|y_{1:i}]$ are of order
$N^{-1/2}$. Previous filtering methods for the diffusion problems
we consider are based on (i) discretising time and introducing $M$
intermediate time points between each observation time; (ii) using
an Euler, or higher order, approximation to the diffusion
(\ref{eq:SDEmult}); and (iii) applying a particle, or other,
filter to this approximate discrete time model. See for example
\cite{Crisan/DelMoral/Lyons:1999}. Results giving the order of the
errors in one such scheme are given by \cite{mora:jaco:prot:2001}.
For Models such as (A) and (B) the errors are of order $N^{-1/2}$
provided that the number of intermediate time steps $M$ between
each observation increases at a rate $N^{1/2}$. Thus for fixed
computational cost $K\propto MN$ the errors decrease at a rate
$K^{-1/3}$. For models such as (C), where the likelihood depends
on the path of the state between two successive observations, the
rate at which errors decrease will be slower, for example
$K^{-1/4}$ \cite[]{mora:jaco:prot:2001}, or $K^{-1/6}$
\cite[Theorem 1.1 of][]{Crisan/DelMoral/Lyons:1999}.


\section{Generalised Poisson Estimators} \label{sec:GPE}

We have already motivated the need for the simulation of a
positive unbiased estimator of
\begin{equation} \label{eq:RN}
\expect\left[ E \right]
\hbox{ where }  E =:\exp \left \{-\int_0^t g(\W_s)\mbox{d}s \right
\},
\end{equation}
where the expectation is taken w.r.t.\@ a $d$-dimensional Brownian
bridge $\W$.  In this section we introduce a methodology for
deriving such estimators, and provide theoretical and simulation
results regarding the variance of the suggested estimators. These
results are of independent interest beyond particle filtering, so
we
present our methodology in a general way, where $g$ is an
arbitrary function assumed only to be continuous on $\real^d$. We
assume that $\W_0=\x$ and $\W_t=\z$, for arbitrary $\x,\z \in
\real^d$ and $t>0$. By the time-homogeneity property of the
Brownian bridge our methodology extends to the case where the
integration limits change to $u$ and $u+t$, for any $u>0$.

\cite{besk:papa:robe:fear:2006} proposed an unbiased
estimator of (\ref{eq:RN}), the {\em Poisson
Estimator}:
\begin{equation}
\label{eq:GPE0} \textrm{PE:} \quad
 e^{(\lambda-c)t}
 \lambda^{-\kappa} \prod_{j=1}^{\kappa} \left [c-g(\W_{_{\psi_j}}) \right ]
 \,;
\end{equation}
$\kappa$ is a Poisson random variable with mean $\lambda t$, the
$\psi_j$s are uniformly distributed on $[0,t]$, and $c \in \real$,
$\lambda>0$ are arbitrary constants. (Here and below we assume
that the empty product, i.e.\@ when $\kappa=0$, takes the value
1.)
%
%
The two main weaknesses of the PE are that it may return negative
estimates and that its variance is not guaranteed to be finite.
Both of these problems are alleviated when $g$ is bounded. 
However this is a very restrictive assumption in our context.
Therefore, here, we introduce a collection of unbiased and
positive 
estimators of (\ref{eq:RN}) which generalise the
PE. The methods we consider
allow $c$ and $\lambda$ depend on $\W$, and permit $\kappa$
to have a general discrete distribution.
Firstly,
we need to be able to simulate
random variables $\low$ and $\up$ with
\begin{equation}
\label{eq:up-low} \low \leq g(\W_s) \leq \up,\quad\textrm{~for
all~} s \in [0,t],
\end{equation}
and to be able to simulate $\W_s$ at any $s$, given the condition
implied by (\ref{eq:up-low}). For unbounded $g$ this is
non-trivial. However, both of these simulations have become
feasible since the introduction of an efficient algorithm in
\cite{besk:papa:robe:fact}. An outline of the construction is
given in Appendix A.

Let $\up$ and $\low$ satisfy (\ref{eq:up-low}) and $\psi_j,j \geq
1$, be a sequence of independent uniform random variables on
$[0,t]$. Then, (\ref{eq:RN}) can be re-expressed as follows,
\begin{equation}
\label{eq:GPE-calc}
\begin{split}
\expect
\left[
e^{-\up t}\exp
\left\{ \int_0^t (\up-g(\W_s))
\textrm{d}s
\right\}
\right]
= \expect
\left[
e^{-\up t}
\sum_{k=0}^\infty \frac{1}{k!} \left (\int_0^t
(\up-g(\W_s)) \textrm{d}s \right )^k
\right]
\\
  =
\expect
\left[
e^{-\up t} \expect
\left[
\sum_{k=0}^\infty
\frac{t^k}{k!}  \prod_{j=1}^k (\up-g(\W_{\psi_j}))  \mid \up,\low
\right]
\right]
\\
 =
\expect
\left[
e^{-\up t}
\frac{t^\kappa}{\kappa!p(\kappa \mid \up,\low)} \prod_{j=1}^\kappa
(\up-g(\W_{\psi_j}))
\right],
\end{split}
%
\end{equation}
where $\kappa$ is a discrete random variable with conditional
probabilities $\prob[\kappa=k \mid \up,\low] = p(k \mid
\up,\low)$.
The second equality in the above argument is obtained using
dominated convergence and Fubini's theorem (which hold by
positivity of the summands).

We can derive various estimators of (\ref{eq:RN}) by specifying
$p(\cdot \mid \up,\low)$. The family of all such estimators will
be called the {\em Generalised Poisson Estimator} (GPE):
\begin{equation}
\label{eq:GPE} \textrm{GPE:} \quad e^{-\up t}
\frac{t^\kappa}{\kappa!p(\kappa \mid \up,\low)} \prod_{j=1}^\kappa
(\up-g(\W_{\psi_j}))\,.
\end{equation}
The following Theorem (proved in Appendix B) gives the optimal
choice  for $p(\cdot \mid \up,\low)$.
\begin{theorem}
\label{th:gpe-var} The conditional second moment of the
Generalised Poisson Estimator given $\up$ and $\low$,  is:
\begin{equation}
\label{eq:GPE-var}  e^{-2\up t } \sum_{k=0}^\infty
\frac{t^{k}}{p(k \mid \up,\low) k!^2}\expect\left [ \left
(\int_0^t (\up-g(\W_s))^2\textrm{d}s \right )^k  \mid \up,\low
\right ]\,.
\end{equation}
If \begin{equation} \label{eq:gpe-cond} \sum_{k=0}^\infty
\frac{t^{k/2}}{k!}\expect\left [ \left (\int_0^t
(\up-g(\W_s))^2\textrm{d}s \right )^k  \mid \up,\low \right
]^{1/2} <\infty\,, \end{equation} then the second moment is
minimised by the choice
\begin{equation} \label{eq:GPEopt}
p(k \mid \up,\low) \propto \frac{t^{k/2}}{k!}\expect\left [ \left
(\int_0^t (\up-g(\W_s))^2\textrm{d}s \right )^k \mid \up,\low
\right ]^{1/2}\,,
\end{equation}
with minimum second moment given by
\begin{equation} \label{eq:GPE-opt-var}  \left ( e^{- \up t} \sum_{k=0}^\infty \frac{t^{k/2}}{k!}\expect\left [ \left
(\int_0^t (\up-g(\W_s))^2\textrm{d}s \right )^k  \mid \up,\low
\right ]^{1/2} \right )^2 <\infty\,,~ \textrm{for almost all}~
\up,\low\,.
\end{equation}

\end{theorem}

\noindent Whilst the right-hand side of (\ref{eq:GPEopt}) cannot be
evaluated analytically, it can guide a suitable choice
of $p(\cdot \mid \up,\low)$. If $\W$ were known, the
optimal proposal is Poisson with mean
\begin{equation}
\label{eq:stoch-rate} 
\lambda_\W:=\left( t \int_0^t (\up-g(\W_s))^2 \mbox{d}s
\right)^{1/2} \,.
\end{equation}
\noindent We will discuss two possible ways that
(\ref{eq:stoch-rate}) can be used to choose a good proposal.

A conservative approach takes $p(\cdot \mid \up,\low)$  to be
Poisson with mean $(\up-\low)t$ (an upper
bound of $\lambda_\W$). 
We call this estimator GPE-1. An advantage of GPE-1 is that its
second moment is bounded above by $\expect[e^{-2 \low t} ]$. Thus,
under mild and explicit conditions on $g$, which are contained in
the following theorem (proved in Appendix C), the variance of the
estimator is guaranteed to be finite.
\begin{theorem}
\label{th:bachelier} A sufficient condition for GPE-1 to have
finite variance is that
$$g(u_1,\ldots,u_d) \geq -\delta \sum_{i=1}^d (1+|u_i|),\quad
\textrm{for all } u_i \in \real, 1 \leq i \leq d,~\delta\geq 0. $$
\end{theorem}

Since $\lambda_\W$ is stochastic, an alternative approach is to
introduce a (exogenous) random mean and assume that $p(\cdot
|\up,\low)$ is Poisson with this random mean. For
tractability we choose the random mean to have a Gamma
distribution, when $p(\cdot \mid \up,\low)$ becomes a
negative-binomial distribution:
\begin{equation}
\label{eq:GPE2}
 \textrm{GPE-2:} \quad e^{-\up t} \frac{t^\kappa \Gamma(\beta) (\beta+\gamma_{_{\W}})^{\beta+\kappa}}{\Gamma(\beta+\kappa) \beta^\beta\gamma_{_{\W}}^\kappa}
 \prod_{j=1}^{\kappa} \left [\up-g(\W_{_{\psi_j}}) \right ]
 \,,
\end{equation}
where $\gamma_{_{\W}}$ and $\beta$ denote the mean and the
dispersion parameter respectively of the negative binomial. Since
the negative-binomial has  heavier tails than the Poisson Estimator, GPE-2
will have finite variance whenever there exists a PE with finite
variance. On the other hand, big efficiency gains can be achieved
if $\gamma_{_{\W}}$ is chosen to be approximately
$\expect[\lambda_\W \mid \up,\low]$. There is a variety of ad-hoc
methods which can provide a rough estimation of this expectation.
Applying Jensen's inequality to exchange the integration with the
square power in (\ref{eq:stoch-rate}), and subsequently
approximating $\expect[\,g(\W_s) \mid \up,\low \,]$ by
$g(\expect[\W_s])$, suggests taking
\begin{equation}
\label{eq:l-g}\gamma_{_{\W}} = t \up - \int_0^t g\left(
\x\frac{t-s}{t}+\y\frac{s}{t} \right)\mbox{d}s>0 \,.
\end{equation}
 A simulation study (part of which is presented in Section
 \ref{sec:sim-study} below)
reveals that this choice works very well in practice and the
GPE-2 has up to several orders of magnitude smaller variance than
the PE or the GPE-1. The integral can usually be easily evaluated,
otherwise a crude approximation can be used.

We have confined our presentation to the case where the
expectation in (\ref{eq:RN}) is w.r.t.\@ the Brownian bridge
measure. Nevertheless, as pointed out in
\cite{besk:papa:robe:fear:2006} the PE can be constructed in
exactly the same way when the expectation is taken w.r.t.\@ an
arbitrary diffusion bridge measure, as long as exact skeletons can
be simulated from this measure. The GPE can also be implemented in
this wider framework, provided that the process $\W$ can be
constructed to satisfy (\ref{eq:up-low}).

\subsection{Simulation study}
\label{sec:sim-study}

%
%


We consider  a smooth bounded test function $g(u) =
(\sin(u)^2+\cos(u)+1)/2$. This has been chosen in view of Example
1. The function $g$ is  periodic,  with period $2 \pi$. In $[0,2\pi]$
it has local minima at 0 and $2\pi$, global minimum at $\pi$ and
maxima at $\pi/3$ and $5\pi /3$. Since $g$ is bounded by $9/8$ we
can construct a PE which returns positive estimates by setting $c
\geq 9/8$. Under this constraint, \cite{besk:papa:robe:fear:2006}
argued that a good choice is $c=\lambda=9/8$. Simulation
experiments suggested that the performance of the GPE-2 is quite
robust to the choice of the dispersion parameter $\beta$. We have
fixed it in our examples to $\beta=10$.  Table \ref{tb:comp}
summarizes estimates of the variance of the estimators based on
$10^4$ simulated values.
\begin{table}
\centering
\begin{tabular}{|ccccc|}
\hline
  & Estimator & $x=0,z=0$ & $x=0,z=\pi$ & $x=\pi,z=\pi$  \\
\hline
variance          & PE & 0.202              & 0.200  & 0.027 \\
                  & GPE-1  &$4.21\times 10^{-3}$& 0.208  & 0.034 \\
                  & GPE-2  &$2.08\times 10^{-3}$& 0.220  & 0.033 \\
                  & Var($E$)  &$3.74\times 10^{-5}$& $3.27\times 10^{-3}$  &$ 4.72\times
10^{-3}$ \\
$\expect[\kappa]$ & PE & 1.118              & 1.126  & 1.121 \\
                  & GPE-1  & 0.130              & 1.091  & 0.744 \\
                  & GPE-2  & 0.119              & 0.329  & 0.735 \\
\hline
\end{tabular} \caption{Monte Carlo estimates of the variance
of four estimators of (\ref{eq:RN}) where $g(u)=
(\sin(u)^2+\cos(u)+1)/2$. For comparison we give also var$(E)$.
 We also report an
estimate of $\expect[\kappa]$. 
We consider three different pairs of starting and ending
points $(x,z)$ and time increment $t=1$. 
The estimates in the
table were obtained from a sample of $10^4$ realisations.}
\label{tb:comp}
\end{table}
%
We see that GPE-2 can be significantly more efficient than PE,
in particular when taking into account $\expect[\kappa]$.
In general, the performance of  PE is  sensitive to
the choice of $c$ and $\lambda$.  GPE-1 is typically less
efficient than  GPE-2. Table 1 also gives the value of
Var$(E)$ which takes significantly smaller values (by
a couple of orders of magnitude) than any of PE, GPE-1 or GPE-2,
illustrating the efficiency cost of these auxiliary variable
constructions in absolute terms.

We have also investigated how the efficiency of the PE and  GPE-2
varies with the time increment $t$ and in particular for small $t$
(results not shown). These empirical results suggest that the
coefficient of variation of the errors of both PE and GPE-2 are
$O(t^\delta)$ for some $\delta>0$; but that the value of $\delta$
differs for the two estimators. In the cases that we investigated,
the GPE-2 appears to have  a faster rate of convergence
than PE.


The results of this simulation study have been verified  for other
functions $g$ (results not shown). We have experimented with  differentiable (e.g.\@
$g(u)=u$) and non-differentiable (e.g.\@ $g(u)=|u|$) unbounded
functions. In these cases it is impossible to design a PE which
returns positive estimates w.p.1. Again, we have found that the
GPE-2 performs significantly better than
the PE. 

It is important to mention that alternative Monte Carlo methods
exist which yield consistent but biased estimates of
(\ref{eq:RN}). One such estimator is obtained by replacing the
time-integral in (\ref{eq:RN}) with a Riemann approximation based
on a number, $M$ say, of intermediate points. This technique is
used to construct a transition density estimator in
\cite{Nicolau:2002} and effectively underlies the transition
density estimator of \cite{durh:gall:2002} (when the diffusion
process has constant diffusion coefficient). The approach of
\cite{durh:gall:2002} has been used in MCMC and filtering
applications
\cite[]{Golightly/Wilkinson:2006,Chib/Pitt/Shephard:2006,ionides}.
In the filtering context it provides an alternative to RWPF, where
the weights are approximated. It is not the purpose of this paper to carry
out a careful comparison of RWPF with such variants. However, as
an illustration we present a very small scale comparison in the
context of estimating the transition density, $p_t(z \mid x)$, of
(\ref{eq:sine-sde}) for $t=1$ and $x,z$ as in Table \ref{tb:comp}.
We compare 4 methods. Two are based on (\ref{eq:td}) and use the
PE and the GPE-2 to generate estimators of the expectation. The
other two, DG-1 and DG-5 are two implementation of the
\cite{durh:gall:2002} estimator, with 1 and 5 respectively
intermediate points. We compare the methods in terms of their root
mean square error divided by the true value (i.e.\@ the
coefficient of variation). As the true value we used the estimate
of the GPE-2. The results of the comparison are presented in Table
\ref{tb:comp2}. Notice that DG-1 and DG-5 simulate many more
variables than GPE-2 to construct their estimates.
\begin{table}
\centering
\begin{tabular}{|cccc|}
\hline
Estimator & $x=0,z=0$ & $x=0,z=\pi$ & $x=\pi,z=\pi$  \\
\hline
PE     & 1.25    & 0.93  & 0.17 \\
GPE-2  & 0.13    & 0.78  & 0.2 \\
DG-1   & 0.5     & 0.45  & 0.3 \\
DG-5   & 0.28    & 0.19  & 0.22 \\
\hline
\end{tabular} \caption{Monte Carlo estimates based on $10^4$
realisations of the root mean square error divided by the true
value of 4 estimators of $p_t(z \mid x)$, of (\ref{eq:sine-sde})
for $t=1$ and various $x,z$. As true value we take the estimate
produced by averaging the estimations given by GPE-2. The number of
intermediate points used for each estimator are 1 and 5 for
DG-1 and DG-5 respectively; the number of Brownian bridge simulations for PE and GPE-2 are given in Table \ref{tb:comp}
($\expect[\kappa]$). } \label{tb:comp2}
\end{table}

\section{Comparison of particle filters on the simulated data} \label{sec:comp}

We now demonstrate the performance of the different particle
filters we have presented on the two examples introduced in
Section \ref{sec:models}.

\subsection{Analysis of the sine diffusion}

\begin{figure}[h]
\begin{center}
\includegraphics[angle=270,width=5in]{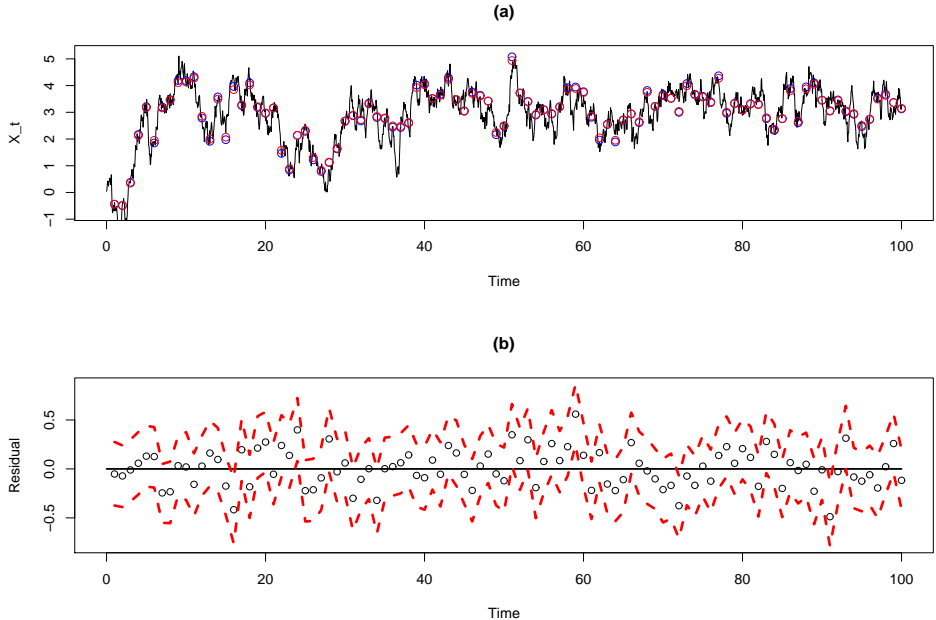}
\caption{\label{Fig:1} Top: A realisation of the sine diffusion
(black line) on $[0,100]$; 100 observations at unit time intervals
(blue circles);
 mean of the filtering distribution of the diffusion at
 the observation times  obtained by
RWPF2 with $N=1,000$ particles (red circles). Bottom: the
difference between observed data and filtered means (circles) and
$90\%$ credible intervals (red dashed lines) from RWPF2. (Whilst
for clarity they are shown for all times, the credible intervals
were only calculated at the observation times.)}
\end{center}
\end{figure}

We first consider analysing the sine diffusion of Example 1. The
simulated data is shown in Figure \ref{Fig:1}(top).
 We compare
four implementations of the particle filter each of which avoids
time-discretisations  by using methodology based on the Exact
Algorithm (EA) for simulating diffusions: i) EPPF, which uses EA
for implementing a bootstrap filter, ii) ESPF, which adapts EA to
simulate by rejection sampling from the filtering densities, iii)
RWPF1, an implementation of RWPF using PE (see Table
\ref{tb:comp}) to simulate the weights, iv) RWPF2, an
implementation of RWPF using GPE-2 to simulate the weights.
Details on the implementation of EPPF and ESPF are given in
Appendix D.

In this simple example ESPF is more efficient than EPPF, since it
has the same computational cost, but it is proposing from the
optimal proposal distribution. However, we have efficiently
implemented  ESPF exploiting several niceties of this simple
model, in particular the Gaussian likelihood and the fact that the
drift is bounded. In more general models implementation of ESPF
can be considerably harder and its comparison with EPPF less
favorable due to smaller acceptance probabilities.

In this context where $\phi$ is bounded one can speed up the
implementation of GPE-2 with practically no loss of efficiency by
replacing $\up$ in (\ref{eq:GPE2}) and (\ref{eq:l-g}) by $9/8$
which is the upper bound of $\phi$. In this case, there is no need
to simulate $\up$ and $\low$. We have implemented this
simplification in the RWPF2.

Algorithms EPPF, RWPF1-2 used the stratified re-sampling algorithm
of \cite{Carpenter/Clifford/Fearnhead:1999}, with re-sampling at
every iteration. For RWPF1-2 we chose the proposal distribution
for the new particles based on the optimal proposal distribution
obtained if the sine diffusion is approximated by the Ozaki
discretisation scheme (details in Appendix E). For EPPF we chose
the $\beta_i^{(k)}$s to be those obtained from this approximation.

The number of particles used in each algorithm was set so that
each filter had comparable CPU cost, which resulted in 500, 500,
910 and 1000 particles used respectively for each algorithm. For
these numbers of particles, EPPF and ESPF on average required the proposal of
1360 particles and required 675 Brownian bridge
simulations within the accept-reject step (iii) at each iteration
of the algorithm. By comparison RWPF1 and RWPF2 simulated
respectively 910 and 1000 particles and required on average 1025
and 850 Brownian bridge simulations to generate the random weights
at each iteration.

Note that the comparative CPU cost of the four algorithms, and in
particular that of EPPF and ESPF as compared to RWPF1-2 depends on
the underlying diffusion path. The acceptance probabilities within
EPPF and ESPF depend on the values of $x_i^{k_{i,j}}$ and
$x_{t_{i+1}}$, and get small when both these values are close to
$0 (\mbox{mod } 2\pi)$. (in the long run the diffusion will visit
these regions infrequently and will stay there for short periods.)
Thus, simulated paths which spent more (or less) time in this
region of the state-space would result in EPPF and ESPF having a
larger (respectively smaller) CPU cost.

\begin{figure}
\begin{center}
\includegraphics[angle=270, width=4in]{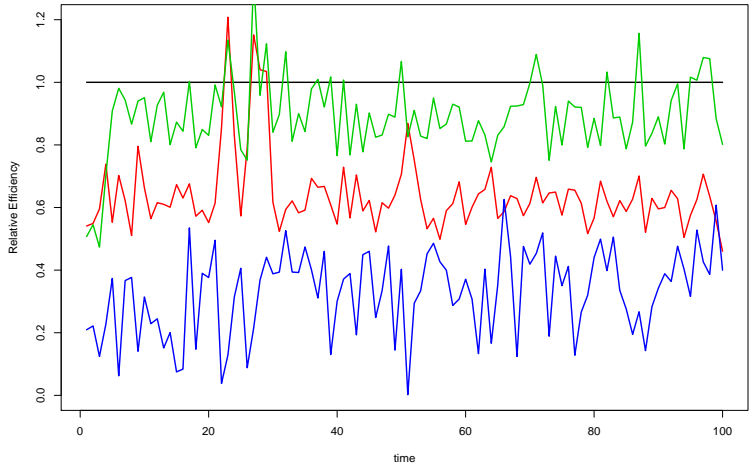}
\caption{\label{Fig:2} Relative efficiency of the 4 particle
filter algorithms at estimating the filtering mean
$\expect[X_{t_i}|y_{1:i}]$. Each line gives the relative
efficiency of one algorithm  compared to RWPF2 (black: RWPF2,
green: RWPF1, red: ESPF, blue: EPPF).  See text for details. 
}
\end{center}
\end{figure}

We compared the four filters based on the variability of estimates
of the mean of the filtering distribution of the state across 500
independent runs of each filter. Results are given in Figure
\ref{Fig:2}, while output from one run of RWPF2 is shown in Figure
\ref{Fig:1}. The comparative results in Figure \ref{Fig:2} are for
estimating the mean of the filtering distribution at each
iteration (similar results were obtained for various quantiles of
the filtering distribution). They show RWPF2 performing best with
an average efficiency gain of $15\%$ over RWPF1, $50\%$ over ESPF
and $200\%$ over EPPF. Interpretation of these results suggest
that (for example) ESPF would be required to run with $N=750$
(taking $1.5$ times the CPU cost for this data set) to obtain
comparable accuracy with RWPF2.

Varying the parameters of the model and implementation of the
algorithms will affect the relative performance of the algorithms.
In particular increasing (or decreasing) $\sigma^2$, the variance
of the measurement error, will increase (respectively decrease)
the relative efficiency of EPPF relative to the other filters.
Similar results occur as $\Delta_i$ is decreased (respectively
increased). The relative performance of the other three algorithms
appears to be more robust to such changes. We considered
implementing EPPF with $\beta_i^{(k)}=w_i^{(k)}$; and also using
an Euler rather than an Ozaki approximation of the sine diffusion
to construct the proposal distribution for RWPF1-2, but neither of
these changes had any noticeable effect on the performance of the
methods. We also considered re-sampling less often, setting
$C=N/4$ in step PF1 of the RWPF algorithm (so re-sampling when the
effective sample size of the $\beta_i^{(j)}$s was less than $N/4$)
and this reduced the performance of the algorithms substantially
(by a factor of 2 for RWPF1-2).



We also investigated the effect of increasing the amount of time, $\Delta$, between observations.
To do this we used the above data taking
(i) every 10th; or (ii) every 20th time-point.

To measure the performance of the filter for these different
scenarios we used the Effective Sample Size (ESS) of
\cite{Carpenter/Clifford/Fearnhead:1999}. ESS is calculated
based on the variance of estimates of posterior means across
independent runs of the filter, but this variance is compared to
the posterior variance to give some measure of how many
independent draws from the posterior would produce estimators
of the same level of accuracy.
We focus on
estimates of the posterior mean of the state at observation times;
and if $s^2$ is the sample variance of the particle filter's
estimate of $\expect[X_{t_i}|y_{1:i}]$ across 100 independent
runs, and $\hat{\sigma}^2$ is an estimate of
$\mbox{Var}[X_{t_i}|y_{1:i}]$, then the ESS is
$\hat{\sigma}^2/s^2$. Note that comparing filters by their ESS is
equivalent to comparing filters based on the variance of the
estimators.

Table \ref{Tab:R1} gives ESS values for the different values of $\Delta$.
We see that the ESS values drops dramatically as $\Delta$ increases, and the filter is inefficient for $\Delta=20$.
This drop in performance is due to the large variability of the random weights in this case. The variability of these weights
is due to (a) the variability of
\begin{equation} \label{eq:r1}
 \exp \left
\{-\int_{t_i}^{t_{i+1}} g(\W_s)\mbox{d}s \right \},
\end{equation}
across different diffusion paths; and (b) the Monte Carlo
variability in estimating this for a given path. To evaluate what
amount is due to (a), we tried a particle filter that estimates
(\ref{eq:r1}) numerically by simulating the Brownian Bridge at a
set of discrete time points (for this example we sampled values
every 1/2 time unit) and then using these to numerically evaluate
the integral. This approach is closely related to the importance
sampling approach of \cite{durh:gall:2002,Nicolau:2002}, see
Section \ref{sec:sim-study}.  The results for this filter are also
given in Table \ref{Tab:R1} (note the ESS values ignore any bias
introduced through this numerical approximation), and we again see
small ESS values, particularly for $\Delta=20$. This filter's
performance is very similar to the RWPF, which
 suggests that the Monte Carlo variability in (b) is a small contributor
 to the poor performance of the RWPF in this case.

Finally we tried introducing pseudo observations at all
integer time-intervals where currently no observation is made. The
RWPF is then run as above, but with no likelihood contribution to the
weight at the time-points where there are these uninformative
observations. The idea is that now $\Delta=1$, so that the variance of
the random weights is well-behaved, but we still have
adaptation of the
path of the diffusion in the unit time-interval prior to an
observation to take account of the information in that
observation. Results are again shown in  Table \ref{Tab:R1}, and the
ESS values are very high (and close to the optimal value, that of the
number of particles, 1000). Note that the computational cost is only
roughly doubled by adding these extra pseudo observations; as the
total computational cost for the simulation of the Brownian bridge is unchanged.  These results are reasonably robust to the choice of how frequently to introduce these uninformative observations (results not shown).

\begin{table}
\begin{center}
\begin{tabular}{l|c|c|c}
Filter &  $\Delta=10$&  $\Delta=20$ \\ \hline
RWPF2    & 73 & 5 \\
Discretisation & 80 & 12 \\
pseudoRWPF2  & 923 & 933
\end{tabular}
\caption{\label{Tab:R1} Comparison of filter's mean ESS values for
  different time intervals between observations ($\Delta$). Results
  are for the Random Weight Particle Filter using GPE-2 (RWPF2), a filter that numerically approximates the weight through discretising the diffusion process (Discretisation), and the RWPF after introducing uninformative observations at unit time intervals (pseudoRWPF2).}
\end{center}
\end{table}

\subsection{Analysis of the Cox process}

\begin{figure}[h]
\begin{center}
\includegraphics[angle=270,width=5in]{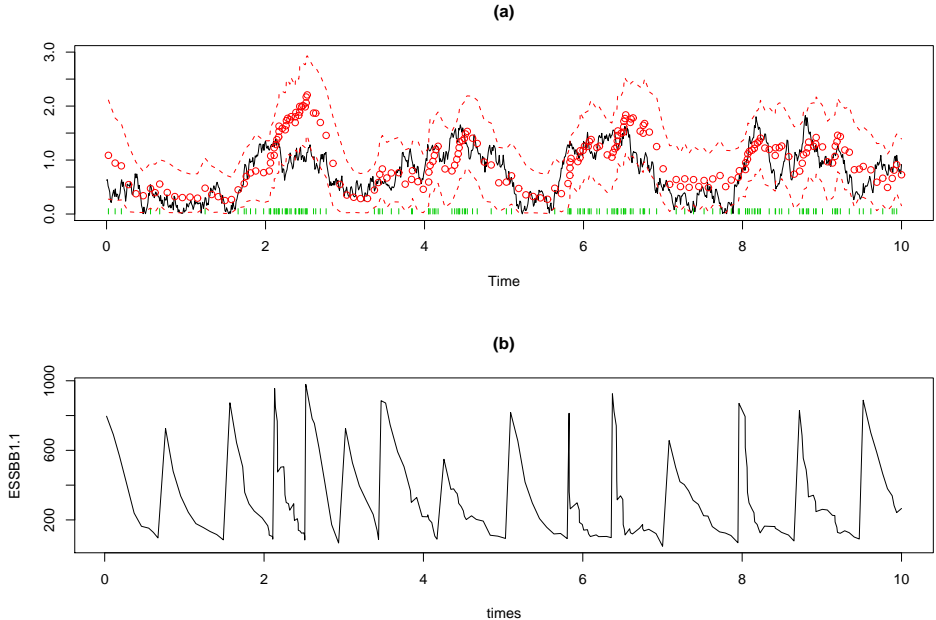}
\caption{\label{Fig:3} Top: Simulation from the Cox process of
Example 2 and results from analysis by the RWPF. The path of the
absolute of the underlying diffusion (black line); observed
arrival times (green dashes); filter estimates from the RWPF (red
circles); and $90\%$ credible interval for the absolute of the
diffusion (red dashed line). (Whilst for clarity they are shown
for all time, the credible intervals  were only calculated at and
apply  for times where filter estimates are shown.) Bottom:
ESS of the RWPFs weights (defined as
$(\sum_{j=1}^Nw_i^{(j)})^2/\sum_{j=1}^N (w_i^{(j)})^2$) over time.  The dramatic increases
in the effective sample sizes correspond to re-sampling times. }
\end{center}
\end{figure}
We now consider applying the random weight particle filter (RWPF)
to Example 2 from Section \ref{sec:models}, the OU-driven Cox
process. The data we analysed is given in Figure \ref{Fig:3}(top).
It is either impossible or difficult to adapt the other two
EA-based particle filters (the EPPF and the ASPF) to this problem.
For instance we cannot implement EPPF as the likelihood function
is not tractable. As such we just focus on the efficiency of the
RWPF in estimating the filtering distribution of  $|X_t|$.

Our implementation of the RWPF was based on proposing particles
from the prior distribution, so $\beta_i^{(k)}=w_i^{(k)}$ and
$q(x_{t_{i+1}}|x_i^{j},y_{i+1})$ is just the OU transition density
$p(x_{t_{i+1}}|x_i^{j})$. We simulated the random weights by
GPE-2. We calculated the filtering density at each observation
time, and also at 56 pseudo-observation times chosen so that the
maximum time difference between two consecutive times for which we
calculated the filtering density was 0.1. This was necessary to
avoid the number of Brownian bridge simulations required to
simulate the weights being too large for long inter-observation
times, and also to control the variance of the random weights (see above).
The likelihood function
for these non-observation times is obtained by removing
$\nu(x_{t_i})$ from (\ref{eq:cox-lik}). 

We set the number of particles to $1,000$ and resampled when the
ESS of the $\beta_i^{(j)}$s was less then 100
($C=N/10$ in step PF1 of the algorithm in Section \ref{PF}).
Whilst results for the sine diffusion suggest that this will
result in an algorithm that re-samples too infrequently, we chose
to have a low threshold so that we could monitor the performance
of the particle filter by how the ESS of the
particle filter weights decay over time. The results of one run of
this filter are shown in Figure \ref{Fig:3}(top). The
computational efficiency of this method can be gauged by Figure
\ref{Fig:3} (bottom) where the ESS of the
$w_i^{(j)}$s is plotted over time.

%
%
%

\section{Discussion}
\label{sec:discuss}

We have described how recent methods for the exact simulation of
diffusions and the unbiased estimation of diffusion exponential
functionals can be used within particle filters, so that the
resulting particle filters avoid the need for time-discretisation.
\comment{ We have considered three such particle filters: (i)
EPPF, which uses the EA to propagate particles according to the
correct system diffusion; (ii) ESPF which uses the idea of the EA
to propagate particles from the optimal proposal distribution; and
(iii) RWPF which implements an auxiliary particle filter, but
simulates the weights that are allocated to each particle. We have
shown that the RWPF is equivalent to an auxiliary particle filter
on an expanded state. We have established a central limit theorem
for all three of these particle filters, the consequence of which
is that each of these filters has a better rate of convergence (in
terms of CPU time) over methods that use time-discretisation. }
Among the approaches we have introduced special attention was
given to  RWPF which implements an auxiliary particle filter, but
simulates the weights that are allocated to each particle. We
showed that this methodology is equivalent to an auxiliary
particle filter applied to appropriately expanded model. We expect
that this methodology will have interesting applications to
different models than those considered in this paper, which
however involve intractable dynamics or likelihoods.

\comment{ Of the three approaches, we advocate the use of the
RWPF. The RWPF has an advantage over EPPF in that it is
straightforward to use information from the observation to inform
the proposal distribution of the particles. The RWPF can also be
more easily adapted to different problems than the EPPF or ESPF,
for example the EPPF cannot be used for the Cox model of Example
2; while as ESPF is based on rejection sampling, as opposed to
importance sampling, it requires the need to bound importance
weights, and will require tight bounds to work efficiently.
Furthermore, even on the Sine model of Example 1, where it is
straightforward to implement ESPF in an efficient manner, we found
that RWPF still gave  superior performance over ESPF. Moreover,
the RWPF is more amenable to software implementation and this is
work in progress. }

We have focused on the filtering problem,
estimating the current state given observations to date. However,
extensions to prediction are trivial -- merely requiring
the ability to simulate from the state equation, which is possible
via the EA algorithms. It is also straightforward to use the idea
of \cite{Kitagawa:1996}, where each particle stores the history
of its trajectory, to get approximations of the smoothing density
(the density of the state at some time in the past given the
observations to date).

Note that while particles store values of the state only for each
observation time, it is straightforward to fill in the
diffusion paths between these times to produce inferences about
the state at any time. A particle approximation to the
distribution of $(\X_s, t_{i-1}<s<t_i)$, conditionally on the data
$y_{1:i}$ can be constructed using the current set of weighted
particles $\{(\x_{i-1}^{(j)},\x_i^{(j)})\}_{j=1}^N$ with weights
$\{w_i^{(j)} \}$, as follows. Firstly we need to introduce some
notation; we denote by $\x_{i-1|i}^{(j)}$ the value of the
particle at time $t_{i-1}$ from which the $j$th particle at time
$t_i$ is descended.
 The
particle approximation is given by a set of weighted paths
$\{(\x_s,t_{i-1}<s<t_i)^{(j)} \}_{j=1}^N$ with weights $\{
w_i^{(j)}\}$. Each path is a diffusion bridge starting from
${\x}_{i-1|i}^{(j)}$ and finishing at $\x_i^{(j)}$ and it can be
simulated using  EA, as described in \cite{besk:papa:robe:2004}
and \cite{besk:papa:robe:fact}. In observation regimes (A) and (B)
the EA is applied to simulate a diffusion bridge with density
w.r.t.\@ the Brownian bridge measure given by
$\exp\{-\int_{t_{i-1}}^{t_i} \phi(\X_s) \mbox{d}s \}$, whereas in
regime (C) the corresponding density is
$\exp\{-\int_{t_{i-1}}^{t_i} (\phi(\X_s)+\nu(\X_s)) \mbox{d}s \}$.
This representation can be directly exploited to draw inferences
for any function of a finite skeleton of $\X$ in-between
observation times.

\subsubsection*{Appendix A: The layered Brownian motion}
\label{appendix:layer}

The algorithm proposed in \cite{besk:papa:robe:fact} starts by
creating a partition of the sample space of $\W$  for the given
$\W_0=\x$ and $\W_t=\y$. Writing $\x=(x_1,\ldots,x_d)$, for a
user-specified constant $a>\sqrt{t/3}$, a sequence of subsets of
$\real^d$ is formed as $A_j = \{\u=(u_1,\ldots,u_d):
\min(x_{t_i},y_i) - ja < u_i \leq \max(x_{t_i},y_i)+ja  \},~j \geq
0$, where $\cup_j A_j = \real^d$. This sequence defines a
partition of the sample space of the form $\cup_{j=1}^\infty D_j$,
where a path belongs to $ D_j$ if and only if the path has
exceeded the bounds determined by $A_{j-1}$ but not the bounds
determined by $A_j$. In \cite{besk:papa:robe:fact} it is shown how
to simulate the random variable which determines which of the
$D_j$s $\W$ belongs to, and how to simulate $\W$ at any collection
of times conditional on this random variable, the
{\em layered Brownian bridge} construction.
Since $g$ is assumed continuous, knowing $W \in D_j$  is sufficient to
determine $\up$ and $\low$ which satisfy (\ref{eq:up-low}).
In fact, in the simplified
setting where $g$ is bounded, as in the sine diffusion of Example
1, the layered Brownian bridge construction can be avoided since
it is easy to choose $\up$ and $\low$ independently of $\W$.

\subsubsection*{Appendix B: Proof of Theorem \ref{th:gpe-var}}
\label{appendix:proof}


\begin{equation*}
I:=  \frac{t^\kappa}{\kappa!p(\kappa \mid \up,\low)}
\prod_{j=1}^\kappa (\up-g(\W_{\psi_j}))\,.
\end{equation*}

Then, (\ref{eq:GPE-var}) is established as follows:
\begin{eqnarray*}
 \expect[I^2
\mid \up,\low] & = &  \expect[\, \expect[I^2 \mid \kappa,\W]\,] =
 \expect \left [ \frac{t^{2\kappa}}{(\kappa! p(\kappa \mid
\up,\low))^2} \left ( \int_0^t \frac{(\up -g(\W_s))^2}{t}\mbox{d}s
\right
)^\kappa \right ] \\
& = &  \expect \left [ \frac{t^\kappa}{(\kappa ! p(k \mid
\up,\low))^2} \expect \left [ \left( \int_0^t (\up
-g(\W_s))^2\mbox{d}s \right )^\kappa \mid \up,\low,\kappa \right ]
\right ] \\
& = &  \sum_{k=0}^\infty \frac{t^{k}}{p(k \mid \up,\low)
k!^2}\expect\left [ \left (\int_0^t (\up-g(\W_s))^2\textrm{d}s
\right )^k  \mid \up,\low \right ]\,.
\end{eqnarray*}
Fubini's theorem and dominated convergence are used above (valid
since the integrands are positive a.s.). (\ref{eq:GPE-opt-var}) is
obtained using the following result (which can be easily proved
using Jensen's inequality). Let $f_i>0$ for $i=1,2,\ldots$. Then
the sequence of $p_i$s which minimize $\sum_{i=0}^\infty f_i/p_i$
under the constraint $\sum p_i =1$ is given by $p_i =
\sqrt{f_i}/\sum \sqrt{f_i}$.

\subsubsection*{Appendix C: Proof of Theorem \ref{th:bachelier}}
GPE-1$\le e^{-\low t}$ so that the result holds if $ \expect[
 e^{-\low t}
] < \infty $, where the expectation  is w.r.t. a  $d$-dimensional
Brownian bridge from ${\bf x}$ at time $0$ to ${\bf y}$ at time
$t$. However
\begin{eqnarray*}
\expect[
 e^{-\low t}
]
& = & \int_0^\infty \prob [e^{-L_{\W }} > w] \mbox{d}w \\
& = & \int_0^\infty \prob [\low < - \log w ] \mbox{d}w
 \le  \int_0^\infty \prob[ \delta \sum_{i=1}^d(1 + M_i) > \log w
] \mbox{d}w
\end{eqnarray*}
where $M_i = \sup_{0 \le s \le t} |W_i|$ using the the growth
bound in Theorem \ref{th:bachelier}. Furthermore,
\begin{eqnarray*}
 \int_0^\infty \prob[
\delta \sum_{i=1}^d(1 + M_i) > \log w ] \mbox{d}w &\le  &
\int_0^\infty \sum_{i=1}^d \prob[ \delta (1 + M_i) > d^{-1} \log w
]
\mbox{d}w \\
& = &  \int_0^\infty \sum_{i=1}^d \prob[  M_i > (d\delta)^{-1}
\log w  - 1] \mbox{d}w \ .
\end{eqnarray*}
It remains therefore to bound the $d$ integrals on the right hand
side of this expression. However from the Bachelier-Levy formula
for hitting times for Brownian motion and bridges,
$$
\prob [M_i > v ] \le  \exp\left \{-2 (v - \max \{x_i, y_i\} )^2/t
\right \} + \exp \left \{-2 (\min \{x_i, y_i + v\} )^2/t\right \}
$$
 and so
\begin{eqnarray*}
\prob[  M_i > (d\delta)^{-1}  \log w  - 1] & \le & \exp\{-2
((d\delta)^{-1}  \log w - 1) - \max \{x_i, y_i\} )^2/t\} \\ & + &
\exp\{-2 (\min \{x_i, y_i + (d\delta)^{-1}  \log w  - 1)\} )^2/t\}
\end{eqnarray*}
which recedes like $w^{- k \log w}$ as $w \to \infty $ thus
concluding the proof.

\subsubsection*{Appendix D: EPPF and ESPF for Example 1}

EPPF generates the new particles according to the following
procedure:
\begin{itemize}
\item[(i)] choose one of the current particles $x_i^{(k_{i,j})}$,
where particle $j$ is chosen w.p.\@ $\beta_i^{(j)}$;
 \item[(ii)] propose $x_{t_{i+1}}$ from Normal with mean
$x_{i}^{(k_{i,j})}$, and variance $\Delta_i$; \item[(iii)]
accept
this proposal w.p. $\exp( -\cos(x_{t_{i+1}})-1)$; if
proposal is rejected return to (i). \item[(iv)] accept this
proposal with probability
\[
\expect\left [ \exp \left \{-\int_0^{\Delta_i}\phi(W_s)\mbox{d}s
\right \}\right],
\]
where expectation is with respect to the law of a Brownian Bridge
from $W_0=x_{i}^{(k_{i,j})}$ and $W_{\Delta_i}=x_{t_{i+1}}$, and
$\phi$ is given in (\ref{eq:phi}). If the proposal is rejected
return to (i), otherwise $x_{t_{i+1}}$ is the new particle at time
$t_{i+1}$ with weight $w_{i+1}^{}= f(y_{i+1}|x_{i+1}^{})$.
\end{itemize}
(iv) is performed using retrospective
sampling as described in \cite{besk:papa:robe:2004}.

ESPF proceeds as above but with steps (i) and (ii) replaced by the
step
\begin{itemize}
\item[(i')] propose $(x_i^{(k_{i,j})}, x_{t_{i+1}})$ according to the
density proportional to
\[
\exp\left\{-\cos(x_{i}^{(k_{i,j})})-\frac{\left
(y_{i+1}-x_{i}^{(k_{i,j})}\right)^2}
{2(\sigma^2+(\Delta_i))}\right\}
\exp\left\{-\frac{(x_{t_{i+1}}-\eta)^2}{2\tau^2}\right\}
\]
where $ \eta=
({\sigma^2+\Delta_i})^{-1} ( {
x_{i}^{(k_{i,j})}\sigma^2+\Delta_iy_{i+1}} ) $, and
$ \tau=\sigma^2\Delta_i/(\sigma^2+\Delta_i). $
\end{itemize}
The 
algorithm is repeated until $N$ values for
$x_{t_{i+1}}$ are accepted,
each with weight $1/N$.

\subsubsection*{Appendix E: Proposal Distribution for Example 1}
\label{appendix:ProposalE1}

Consider a diffusion satisfying SDE (\ref{eq:SDEmult}), with $d=1$
for simplicity. The Ozaki approximation of this SDE is based on a
first order Taylor
expansion of the drift about some value $x$. 
For the sine diffusion of Example 1, we get the following
approximating SDE
\[
 \mathrm{d}\tilde{X}_s = -\cos(x)[x-\tan(x)-\tilde{X}_s]\,
\mathrm{d}s + \mathrm{d}B_s.
\]
So $\tilde{X}_s-(x-\tan(x))$ is an OU process as defined in
Example 2 
with $\rho=\cos(x)$ and $\sigma = 1$.
To calculate $q(x_{t_{i+1}}|x_i^{(j)},y_{i+1})$ we compute
the product of the transition density given by the Ozaki approximation
about $x=x_i^{(j)}$ and the likelihood function
$f(y_{i+1}|x_{t_{i+1}})$. Defining
$\tau^2=(1-\exp\{-2\cos(x_i^{(j)})\Delta_i\})/(2\cos(x_i^{(j)}))$,
and
$\eta=x_i^{(j)}-\tan(x_i^{(j)})(1-\exp\{-\cos(x_i^{(j)})\Delta_i\})$
we get that $q(x_{t_{i+1}}|x_i^{(j)},y_{i+1})$ is Normal
with mean $(\eta\sigma^2+y_{i+1}\tau^2)/(\tau^2\sigma^2)$ and
variance $\eta^2\tau^2/(\eta^2+\tau^2)$.
Furthermore we calculate
$
\beta_i^{(j)} \propto w_i^{(j)}
\mathcal{N}_{\tau^2+\sigma^2}(y_{i+1}-\eta).
$

\subsubsection*{Appendix F: Central Limit Theorem}
\label{appendix:CLT}

For notational simplicity, we consider a special case of our
particle filter, chosen to resemble those considered in
\cite{Chopin:2004}. We choose our proposal
density for time $t_{i+1}$ to have $\beta_j=w^{(j)}_i$; and we
assume iid sampling of $X_{t_i}^{(j)}$ in step PF1.  The particle
filter of \cite{Chopin:2004} splits up simulating particles at
time $t_{i+1}$ into (i) a resampling of particles at time $t_i$;
and (ii) a propagation of each of these particles to time
$t_{i+1}$. Our assumption of iid sampling is equivalent to the
multinomial resampling case of \cite{Chopin:2004}. (The conditions
for the central limit theorem are the same if the residual
sampling methods of \cite{liu:chen:1998}, but the variances
differ.) For simplicity we consider
observation model (A) or (B),
though the result extends easily to observation
model (C).

Let
$\theta_{i}^{(j)}=(x_{t_i}^{(j)},x_{t_{i-1}}^{(k_{i,j})})$, where
$k_{i,j}$ is the index sampled in step PF1 when simulating the $j$
particle at time $t_i$ and $\theta_{i}^{(j)}$ is the $j$th particle
at time $t_i$ together with the particle at time $t_{i-1}$ from
which it is descended. 
Also let $\expect_{\theta
_i}$ denote  conditional expectation given $\theta _i$. Similarly,
let $\mu_i(\theta_i)=\mu_g(x_{i-1},x_i,t_{i-1},t_i)$, and denote
by $R_{i}$ the unbiased estimator of $\mu_i(\theta_i)$, i.e.\@
$\expect[R_{i}]=\mu_i(\theta _i)$. An important
quantity is $\sigma _i^2(\theta _i)= \mbox{Var}(R_{i})$.

We define $\expect_i[\varphi]$ and $\mbox{Var}_i(\varphi)$ to be the
posterior mean and variance of an arbitrary function $\varphi(\theta)$ at
time $i$, and
consider Particle Filter estimates of  $\expect_i[\varphi]$.
Let $\tilde{\pi}_i(\theta_i)$ be
the density $p(x_{t_{i-1}}|y_{1:i-1})q(x_{t_i}|x_{t_{i-1}})$.
Finally define $\expect_{q_i}[\varphi]$ and $\mbox{Var}_{q_i}(\varphi)$ to be
shorthand for the conditional expectation and variance of $\varphi(\theta_i)$
with respect to $q(x_{t_i}|x_{t_{i-1}})$ (which are functions of $x_{t_{i-1}}$).
We denote
$\|\cdot \|$ to be the Euclidean norm and define recursively $\Phi_i$ to
be the set of measurable functions $\varphi$ such that for some $\delta>0$
$
\expect_{\tilde{\pi}_i}[ \| h_i R_i\varphi \|^{2+\delta}]<\infty
$,
and that the function $x_{t_{i-1}}\mapsto\expect_{q_i}[h_i \mu_i\varphi]$ is
in $\Phi_{i-1}$.

\begin{theorem} \label{thm:CLT}
Consider a function $\varphi$; define
$\tilde{V}_0=\mbox{Var}_{\tilde{\pi}(x_0)}(\varphi)$, and by induction:
\begin{equation} \label{CLT1}
\tilde{V}_i(\varphi)=\hat{V}_{i-1}\left\{\expect_{q_i}[\varphi]\right\}+
\expect_{i-1}\left\{\mbox{Var}_{q_i}(\varphi)\right\}, \mbox{ for $i>0$,}
\end{equation}
\begin{equation} \label{CLT2}
{V}_i(\varphi)={\tilde{V}_i\left\{\mu_i h_i \cdot(\varphi-\expect_i[\varphi])
\right\}+\mbox{E}_{\tilde{\pi}_i}((\varphi-\expect_i[\varphi])^2\sigma _i^2h_i^2) \over
\mbox{E}_{\tilde{\pi}_i}(\mu_i h_i)^2},
\mbox{ for $i\geq0$},
\end{equation}
\begin{equation} \label{CLT3}
\hat{V}_i(\varphi)=V_i(\phi)+\mbox{Var}_i(\phi), \mbox{ for $i\geq0$}.
\end{equation}
Then if for all $i$ (C1) $x_{t_i}\mapsto 1$ belongs to $\Phi_i$;
(C2) $\expect_{\tilde{\pi}_i}[h_i^2\sigma _i^2]<\infty$; and
(C3) $\expect_{\tilde{\pi}_i}[\sigma _i\varphi h_i]^{2+\delta}<\infty$
for some $\delta>0$; then for any $\varphi \in \Phi_i$, $\expect_i[\varphi]$
and $V_i(\varphi)$ are finite and we have the following convergence in
distribution as the number of particles, $N$, tends to infinity:
\[
N^{1/2}\left\{\frac{\sum_{j=1}^N w_i^{(j)} \varphi(x_{t_i}^{(j)})}
{\sum_{j=1}^N w_i^{(j)}}
-\expect_i[\varphi]\right\}\rightarrow \mathcal{N}(0,V_i(\varphi))
\]
\end{theorem}
\noindent
{\bf Comment}
Equations (\ref{CLT1})--(\ref{CLT3}) refer to the changes in variance of the
weighted particles due to the propagation, weighting and resampling stages at
iteration $i$. Only (\ref{CLT2}) differs from the respective result in
\cite{Chopin:2004}, and this is due to the second term on the
right-hand side, which
represents the increase in variance due to the randomness of the weights.
Condition C1 is taken from \cite{Chopin:2004} and applies to standard particle
filters; conditions C2 and C3 are new and are conditions bounding the variance
of the random weights which ensures that $V_i(\varphi)$ is finite.

\proof We adapt the induction proof in \cite{Chopin:2004},
considering in turn the propagation, weighting and resampling
steps
Our filter differs from the standard
particle filter only in terms of the weighting step; and therefore
we need only to adapt the result of Lemma A2 in
\cite{Chopin:2004}. In fact, (\ref{CLT1}) and (\ref{CLT3}) are
identical to the corresponding quantities in \cite{Chopin:2004},
therefore it remains to show (\ref{CLT2}).
 We define the constant $K=\expect_{\tilde{\pi}_i} [R_i h_i]$ and $\varphi^*=R_i h_i (\varphi - E_i(\varphi) )/K$. Within the enlarged signal space framework, we can apply
Equation (4) of \cite{Chopin:2004}, to give:
\begin{eqnarray*}
V_i(\varphi) &=& {\tilde V}_i \left( \varphi^* \right)
=\hat{V}_{i-1}\left\{\expect_{q_i}[\varphi^*]\right\}+
\expect_{i-1}\left\{\mbox{Var}_{q_i}(\varphi^*)\right\}.
\end{eqnarray*}
Now we can calculate $\expect_{q_i}[\varphi^*]$ by first taking expection over the auxiliary variables (conditional on $\theta_i$). This gives  $\expect_{q_i}[\varphi^*]=\expect_{q_i}[\mu_ih_i(\varphi - E_i(\varphi) )/K]$. Similarly we get
\begin{eqnarray}
\mbox{Var}_{q_i}(\varphi^*)&=&\mbox{Var}_{q_i}(\expect[R_ih_i(\varphi - E_i(\varphi) )/K])+\expect_{q_i}(\mbox{Var}[R_ih_i(\varphi - E_i(\varphi) )/K]) \label{eq:F1} \\
&=& \mbox{Var}_{q_i}(\mu_ih_i(\varphi - E_i(\varphi) )/K)+\expect_{q_i}(\sigma^2_ih^2_i(\varphi - E_i(\varphi) )^2/K^2).
\end{eqnarray}
(Here the expectation and variance in (\ref{eq:F1}) are w.r.t. the auxiliary variables).
Combining these results gives (\ref{CLT2}).
\comment{
\begin{eqnarray*}
V_i &=& {\tilde V}_i \left( {R_i h_i \over {\expect}_i [R_i h_i]} (\varphi - E_i(\varphi) )
\right)
= {\expect}_{{\tilde \pi} _i}
 \left[ {R_i^2 h_i^2 \over {\expect}_i (R_i h_i)^2} (\varphi - E_i(\varphi ) )
 ^2 \right] \\
&=&  {\expect}_{{\tilde \pi} _i}\left[ {\expect}_i\left(  {R_i^2 h_i^2 \over
{\expect}_{{\tilde \pi} _i} (\mu_i h_i)^2} (\varphi -
E_i(\varphi ))
 ^2\right) | \theta_i\right]
=  {\expect}_{{\tilde \pi} _i} \left[
{(\sigma _i^2 (\theta _i) + \mu _i^2) h_i^2 (\varphi - E_i(\varphi ))^2
\over
\expect_{{\tilde \pi} _i} (\mu_i h_i)^2
}
\right]
\end{eqnarray*}
as in (\ref{CLT2}).} The regularity conditions (C1) - (C3)
translate directly also.
\qed

\bibliographystyle{rss}
\bibliography{NonLinPF}
\end{document}